\documentstyle[11pt,newpasp,twoside,graphicx]{article}
\font\klein=cmr10

\def\edcomment#1{\iffalse\marginpar{\raggedright\sl#1\/}\else\relax\fi}
\reversemarginpar

\begin{document}
\title{Parallization of  Stellar Atmosphere Codes}
 \author{Peter H\"oflich}
\affil{Dept. of Astronomy, University of Texas, Austin, TX 78712, USA}

\begin{abstract}
Parallel computing has turned  out to be the enabling technology to solve
complex physical systems. However, the transition from shared memory, 
vector computers to massively parallel, distributed memory systems and,
recently, to hybrid systems poses new challenges to the scientist.
 We want to present a cook-book (with a  very strong, personal bias) based 
 on our experience with parallization of our existing codes. Some of the general tools
 and communication libraries are discussed. 
 Our approach includes a mixture of algorithm, domain and
physical module based parallization. The advantages, scalability and limitations 
of each are discussed at some examples.
 We want show that it becomes easier to write parallel code with increasing
complexity of the physical problem making stellar atmosphere codes beyond the classical
assumptions very suitable. 
\end{abstract}

\section{Introduction}
The last few years saw a shift in high performance computing
from the paradigm of vector computing  to  massively parallel systems
which allow  to solve increasingly complex, physical problems.
 We want to concentrate on general concepts to write parallel code  which may
be helpful to get started and to estimate/choose the {\sl right(!)} number
of CPUs for a given numerical problem.   Due to the rapid advances
in the hardware, any number for the performance or the hardware
mentioned (e.g. Myrinet, Gbit-Ethernet) will  be outdated
long before this contribution goes into print.
 To judge the performance of parallel codes, the relevant quantity is the
ratio between communication and CPU-speed and, thus, the numbers may still be
useful. 

 Ideally, code for massively parallel computers should be designed and
written from scratch, and it should not be based on any legacy code.
 However, this approach is both costly and very time consuming.
 Development cycles may be  longer than changes in the computational landscape, and
the flexibility may be reduced to answer problems in astronomy.
 Sometimes, it may be beneficial to adopt a current code base because of   
faster development cycles, higher flexibility  and, more important, to use  well tested
codes. Here, we want to report our attempt of the latter approach to demonstrate problems
and possible solutions. Most of our code is written in FORTRAN with some C routines.
Our  HYDdrodynamical RAdiation transport code is based  on a number of individual programs which
  have been used to carry out many of the 
analyzes of SNIa and Core Collapse Supernovae ({\cite{H88}, ...,
\cite{H95}, \cite{HHWW01}, ...}).  All components have been  written or adopted
to a  modular form with well defined interfaces which allows
an easy coupling (see Fig.  1) and code verification  by exchanging 
 modules.
 The modules consist of physical units to provide a solution for e.g. 
the nuclear network, the statistical equations to determine the atomic level
population, equation of states, the  opacities, 
the hydro or the  radiation transport problem.
The individual modules are coupled explicitly.
 Consistency between the solutions is achieved iteratively
by perturbation methods  (see H\"oflich 2002, this volume).
 \begin{figure}[ht]
 \vskip -0.5cm
 \hskip 0.5cm \includegraphics[width=8.9cm,angle=270]{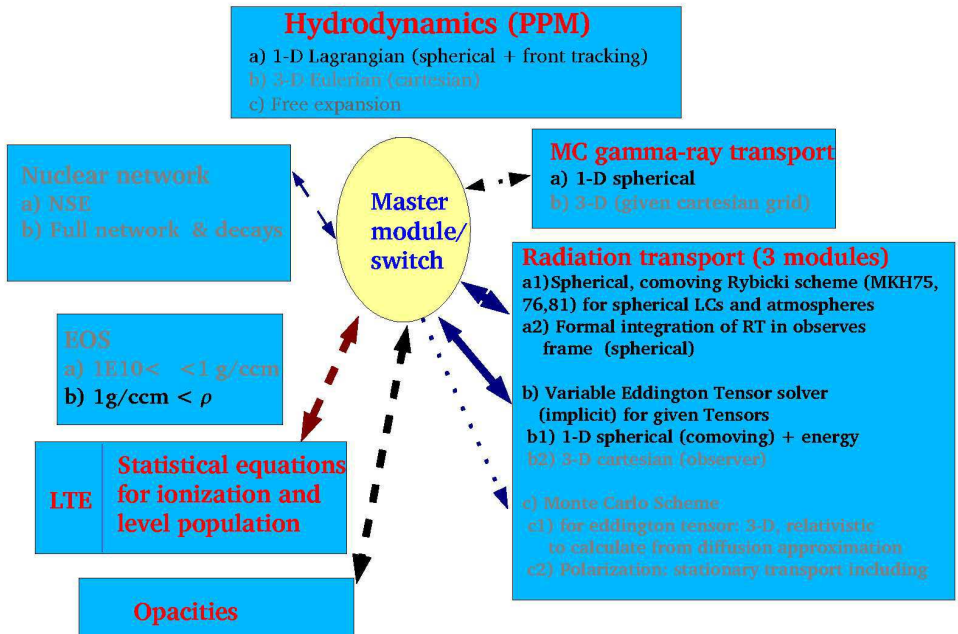}
 \vskip -0.02cm
 \caption{
 Block diagram of our  hydrodynamical radiation transport code (HYDRA) which includes detailed equation of
 states, nuclear and atomic networks.
  The  modules are high-lightened which are used to calculate NLTE-light curves of Type Ia Supernovae.
  Parallization is done on the level of physical modules using 
  the spatial coordinates (dashed), photon
 packages (dotted) and
 frequencies (solid). Within the modules {\sl Opacities} and {\sl Statistical Equations},
 sub-slaves are created to distribute the load according to  line transitions and elements, respectively.
 }
 \label{module}
 \end{figure}

 The goal of this paper is to provide some help in the transition from existing
scalar or vector  to  parallel codes. After some basics and the general concept,
we will discuss various approaches to create parallel code, available tools, and limitations of our
approach.

 The terms  {\sl scalability} and {\sl efficiency} characterize
how well the execution time decrease with the number of processors N(CPU) and the effective speed per processor
compared to N(CPU) one-processor systems.
 The actual numbers have been obtained using a 20 node PC-cluster with dual-Pentium IIs with 400Mhz,
512MB per node and fast Ethernet interconnections. In addition, we used a  16 node  IBM-SP3 system
with 4 CPUs and 1GB per node, interconnected by a  cross-bar.

\section{Basics}

 Three basic approaches to computational resources can be distinguished namely
shared memory systems (e.g. Cray T90, NEC), distributed memory systems 
(Beowulf-clusters) and distributed systems (Grid-computing over the Internet) which
represent a sequence from configurations with highly to very loosely coupled
CPUs  corresponding the trend from  expensive to inexpensive computers.
 Main stream computing uses either shared or distributed
memory systems, and we will concentrate on those.
 Recently, hybrid systems have been introduced which use multiple
CPU-nodes to combine the advantages of both. 

 The basic limitations for  multiple-CPUs is  the communication
namely its {\sl latency} and {\sl bandwidth}. The former measures the time it takes
to establish a communication link between CPUs, the latter provides a measure for the
amount  of data per time  which can be transfered.

 In the following, we want to discuss the concepts and tools.
The strategy for parallization depends on the kind of numerical problem.
 We use either an algorithm based approach, subdivide the computational domain,
 or use a module based approach.

Before discussing details we want to refer to some WEB-sides which we found useful to get started:
\smallskip

\noindent
{\bf \verb"www.links2go.com/topic/parallel_computing"} provides links to a wide variety of  resources.

\smallskip

\noindent
{\bf wotug.ukc.ac.uk/parallel} and  {\bf www.cs.rit.edu/~ncs/parallel.html} are archives for information on  parallel computing.

\smallskip

\noindent
{\bf www.openmp.org} and {\bf www.crpc.rihce.edu/HPFF/home.html} provide a  link to  communications tools for shared memory systems,
namely for OpenMP and HighPerformanceFortran, respectively.

\smallskip

\noindent
{\bf www.mcs.anl.gov} and
{\bf \verb"www.epm.ornl.gov/pvm_home.html"} provide a link to communication tools for distributed memory systems, namely
 MPI and PVM.

\smallskip

\noindent
{\bf www.cs.utk.edu/~browne/perftools-review}   provides a good overview about tools to   analyze MPI and PVM codes.

\smallskip

\noindent
{\bf www.netlib.org}   provides an excellent resource to  math-libraries optimized  for parallel machines including
 ScaLAPACK, BLACS, Linpack, clapack, slatec, sparce, etc.

\bigskip

\subsection{Shared and distributed memory}

The two basic paradigms/concepts of parallel computing are the {\sl shared } and
{\sl distributed } memory model. In shared memory systems,
all  processors have direct access to a unified
memory using high speed buses. This approach avoids the need for
a software based data exchange. On the downside, though, the number of
processors for a computer system using a shared memory is rather small,
and the stringent requirements on the bus-system result in very expensive
computers.  The distributed  memory system allows to connect an unlimited
number of nodes to gain large computing power at a low cost using standard network
technology or relatively inexpensive and, comparably, slow options such as Ethernet,
cross-bars or Myrinet. Each processor has direct access only to the memory
on the local node.  Data communication has to be handled explicitly by the user.
This puts the burden on the user to manage rather sparse resources but
gives more control to  the user.

\subsection{Communication tools}

 In shared memory systems, the  communication channels are sufficiently fast
 to allow parallization for 'moderate' number of CPUs on a loop-level,
very similar to 'classical' vector computers.
 Parallization can be achieved by compiler options (auto-tasking, or HPF)
or, more efficiently, by directives inserted into the code by the user,
e.g. by using  OpenMP (OpenMP Review Board, 2000).
 In distributed memory systems, data communication is achieved by 
explicit message passing using specialized communication libraries, namely
 PVM (Parallel Virtual Machine) (Geist et al. 1994),  the Message Passing Interface (MPI)
(e.g. MPI, 1994, \cite{snir95}), or the new standard MPI2 (\cite{mpi2}).

 PVM has been designed at the Oak Ridge National Laboratory to connect
inhomogeneous clusters of workstations. PVM supports fault tolerance  
and includes a notification scheme, dynamic hosts and it is very
portable due to its PD implementation. Dynamical host allocation means that the
number of tasks can be changed during a run. This is particular useful in complex calculations
to adjust the number of parallel tasks to specific problems, and it may be useful on multi-processor
systems used by a large number of users because it allows to free resources \footnote{Of course,
the maximum number of tasks per user must be limited}. A further advantage of dynamical host allocation
is its "fail-save" feature. If a task terminates due to
a failing hardware  or a numerical problem, a new task can be started
on another CPU or another algorithm to continue the calculation.
 MPI has been developed to provide fast communication on homogeneous
clusters, and it has been adopted as the standard for parallel computing (e.g. Snir et al. 1995).
 Tasks are static but, as for PVM, the resulting codes are
highly portable due to excellent PD implementations (MPICH, 2002; LAM, Byrnes et al. 2002).
 Static task allocation implies that the number of CPU's has to be requested
at the beginning and cannot be adjusted 'on the run' and that the entire job
terminates if a task fails.
 MPI is an excellent choice for domain-based parallization and, numerically, very stable
algorithm as commonly used in hydro calculations whereas PVM has its virtue if
elaborated iteration schemes (e.g. ALI) are used to solve complex systems of non-linear
integro-differential equations such as radiation-hydrodynamical problems including statistical
equations which, sometimes, fail to converge (see H\"oflich, this volume)
\footnote {Note that dynamical task allocation is  not critical if we
calculate NLTE-atmospheres for a few snapshots in time (e.g. Hausschildt, this volume).}. More precisely,
a more aggressive  strategy can increase the convergence rates dramatically but, often, on the
expense of stability.
Even a failor-rate of 1 \% of ALI becomes a  problem  in LC-calculations
which require several thousands of time steps (see below).
 In conclusion, both MPI and PVM are targeted towards distributed memory systems but
their design concepts  are very different.

 To overcome the
problems created by  two separated communication libraries, the new MPI2 standard has been
defined in 1997 by the MPI2-Forum. This new standard combines both the advantages of MPI and
PVM, and it includes important enhancements such as single sided communication which greatly
decrease the latency  for small messages and, thus, boost the overall scalability.
 Unfortunately, no reference implementation
has been released \footnote{The  LAM project is being extended to provide
 a full implementation of MPI2
including dynamical task control}  which hurts portability of codes. Moreover,
most vendor-based implementations are restricted to subset of MPI2 without
dynamical task allocation and single sided communication (e.g. IBM).

 We use a combination of PVM and OpenMP but hope to migrate to MPI2 in the
near future. Using PVM, our code is parallized on the level of large modules, e.g. for
the equation of state, statistical equation, or packages of photons for
radiation transport reducing       the communication overhead.
 To increase the scalability, OpenMP is  employed to
distribute the computations  on the  loop level within time consuming
routines and to take advantage of the on-board shared memory of multi-processor nodes.

\begin{table}[htb]
\caption{Limitations for distributed memory machines due to the communication  latency and bandwidth.
The 'break-even point' is defined by the situation in  which an equal amount of  time is spent
for computations and communications, namely the
the number of floating point operations (FLOPS) per communication and the number of 
in local vs. global memory accesses per communication (mac).
 }
\begin{tabular}{cccccc}
\hline
{CPU-speed} &  {Latency} & {Bandwidth} & {Break-even point} &  {Example} \\
\hline
100 MFlops & $1000 \mu sec$ & $\leq 100 MB/sec $ & 1E5 fpc / 10 mac & Grid-computing \\
100 MFlops & $100 \mu sec$ & $\leq 100 MB/sec $ & 1E4 fpc / 10 mac & Beowulf cluster\\
1GFlops & $1-2 \mu sec$ & $\approx 1...2 GB/sec $ & 1E3 fpc / 10 mac & IBM-SP3 \\
\hline
\end{tabular}
\label{table1}
\end{table}
\subsection{Latency and Bandwidth}

 In theory, multi-processing on N processors may reduce the execution time of a code
by a factor of N (or more). However, in reality, the typical gain is much lower and,
sometimes, multiple CPUs may even cause an increase of the execution time because
 the speed of non-parallel parts stays the same but communication and administrative
overhead increases with the number of processors.

 Distributed memory systems have a large latency and low communication
bandwidth compared to the processing speed of individual CPUs.
 It may be useful to ask for the 'break-even-point' BEP when the communication between two processors requires
as much time as the calculations.  Some typical examples are given in Table 1. Obviously,
parallization is only useful if, per communication, about $10^3$ to $10^5$ floating point operations
are performed and about 10 times more local than global data are accessed. In reality,
an increase in speed by parallization  can be expected if the workload per task is significantly
larger ($\approx number~ of~ processors~ \times ~break~even~point$). Parallization is very inefficient
 on an inner-loop level
 for  all but very  large vectors,

\subsection{ Dynamical Load Balancing}
For both the domain and  (physical) module based parallization,
we distribute the work by dividing the computational load NT-times according to the spatial or frequency
coordinates, use groups of  photons or individual elements.

 In a direct approach, one may distribute packages of size $N = NT/N(CPU)$ as  following:

\bigskip

{\klein do i= 1,N

SEND PACKAGE OF SIZE N TO TASK I

enddo

COLLECT THE RESULTS AND WAIT TILL THE LAST TASK IS FINISHED}

\bigskip

This approach works reasonable well for domain-based parallization if the number of cells are about
the same in all sub-domains (see section 3.2).
 However, it  becomes very inefficient when a complex system of equations has to be solved. Examples are
the solution of implicit equations such as nuclear and atomic networks, or radiation transport by 
Monte-Carlo type methods for which the actual work depends critical  on the optical depth.
 In our example, all processes have to wait for the slowest task.
The direct approach must be modified, and the {\sl dynamical range} DR must be taken into account.
 We define the DR as the ratio  between maximum to minimum CPU time required for a  task.
 In practice, we use package sizes of $\tilde N = NT/(N(CPU)*DR)$. DR must be estimated and
optimization of the code means improving the estimator for DR
\footnote{DR is given as an input parameter but we plan to implement a dynamical estimator.}.
 Small packages increase the overhead in the communication (see table 1).
The package sizes and the corresponding  work loads are reduced with increasing number
of CPUs. Eventually, this limits the scalability of the problem. Often, the performance levels out
at 20 to 50 CPUs on our Beowulf cluster.
For improvements, we employ auto-tasking or OpenMP for parallel algorithms and secondary masters 
on multi-processor nodes with share  memory (see below).

\subsection{Concepts for Parallization}

 Programs may use the Master/Slave approach and/or symmetric multi-processing (Fig. 2).
In the master/slave configurations,
a master-process is used as hub for the communication between different 
'slave' processes and  to balance the load, i.e. to distribute the work. In addition, the
master may be used for sequential calculations during its idle time.
  A   centralized approach makes it
easy to keep track of the timing and  synchronization of tasks, data coherency etc.
 However, the entire communication load is handled by one master process running on a specific
CPU. This  produces a bottle-neck which limits the scalability of the code.
 Direct communication between equal slaves, the  symmetric multi-tasking,
avoids this problem and  allows to take full advantage  of the available bandwidth of the system.
However, sequential calculations have to be done by each slave, and
data coherency may be a problem. In practice, we use a hybrid approach based on a
tree scheme (Fig. 2).
\begin{figure}[ht]
\vskip -3.04cm
\includegraphics[width=4.8cm,angle=270]{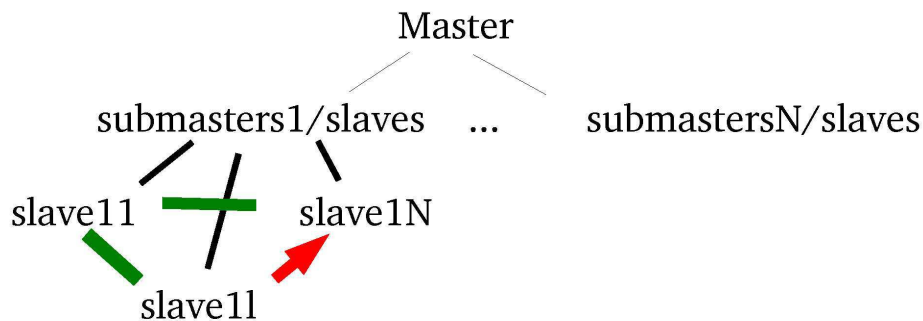}
\vskip +2.194cm
\caption{The 'classical' master/slave approach causes a  bottle
 neck because all communication is  handled by the master-node.
 This limits the scalability. The use of slaves as
 sub-masters and direct communication between slaves reduces this problem. }
\label{slave}
\end{figure}

\section{Classes of Computational Problems}

 The most effective way to parallize code depends largely on the kind of 
numerical problem. In the following, we want to describe some concepts used
in our code HYDRA (H\"oflich, this volume) and show some results for illustration.
 Basically we distinguish between three kind of numerical problems:
1) Algorithm, 2) domain and 3) (physical) module based parallization.

\subsection{Algorithm based parallization}

 During the last few years, very sophisticated  mathematical algorithm have been developed
specially designed for parallel systems, in particular for linear algebra problems, Fourier transformations
etc. . We use this approach for matrix operations and to solve linear equations.
 An effective implementation requires a rewrite of the corresponding routines, extensive testing/optimization
and, often, is very hardware dependent.
 Therefore, we use subroutines available on the NET namely ScaLAPACK, a parallel implementation of
LAPACK.
\begin{figure}[ht]
\includegraphics[width=8.5cm,angle=270]{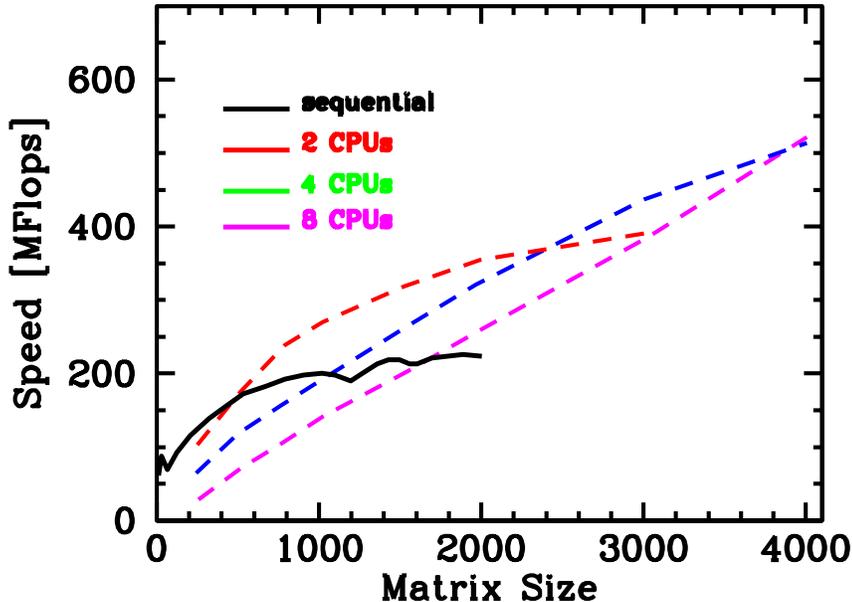}
\vskip -0.04cm
\caption{
Scaling with the number of CPUs for  matrix operation (BLAS routines) on a Linux PC cluster
using MPICH (adopted from http://www.netlib.org).}
\label{matrix}
\end{figure}

 As basic building blocks, ScaLAPACK uses BasicLinearAlgebraSubroutines (BLAS).
 In Fig. 3, the execution speed is given as a function of the rank of a  matrix. Graphs are given
 for a single CPU, 2,4 and 8 processor  PC-clusters with fast Ethernet interconnections using TCP-stacks
or  directly addressing the memory on each node (GAMMA).
 Obviously, there is no gain in speed for a rank $\leq 1000$. 2 CPUs are faster
than 4 CPUs for $N < 2500$. Our tests on an IBM-SP3 showed similar results for
a matrix of rank 2000. By increasing the number of processors by a factor of 16 ($ 4 \rightarrow 64$ processors)
the speedup was  very moderate (factor 2.2).
Therefore, we do not use algorithm based parallization on PC-clusters or distributed
memory systems but restrict this approach  to multi-processor, shared memory nodes.

When using the distributed memory model (PVM or MPI) on
 a dual-PII-node and matrices of rank 100 and 1000, we achieve a speedup of
0.9 and 1.6, respectively.  Using
OpenMP, the corresponding numbers are 1.3 and 1.5.
Therefore, OpenMP is employed  for ranks $\leq 1000 $ and, for  larger matrices, we use PVM or MPI.

As a general rule  for linear algebra problems with of moderate rank ($\leq 5000$),
 parallization should be restricted  to shared memory systems/nodes.
 Optimized algorithm and machine  specific libraries should be used.
 For small matrices ($N \leq 1000$), OpenMP seems to give better performance
than explicit message parsing.
  
\begin{figure}[ht]
\hskip 0.5cm \includegraphics[width=9.5cm,angle=270]{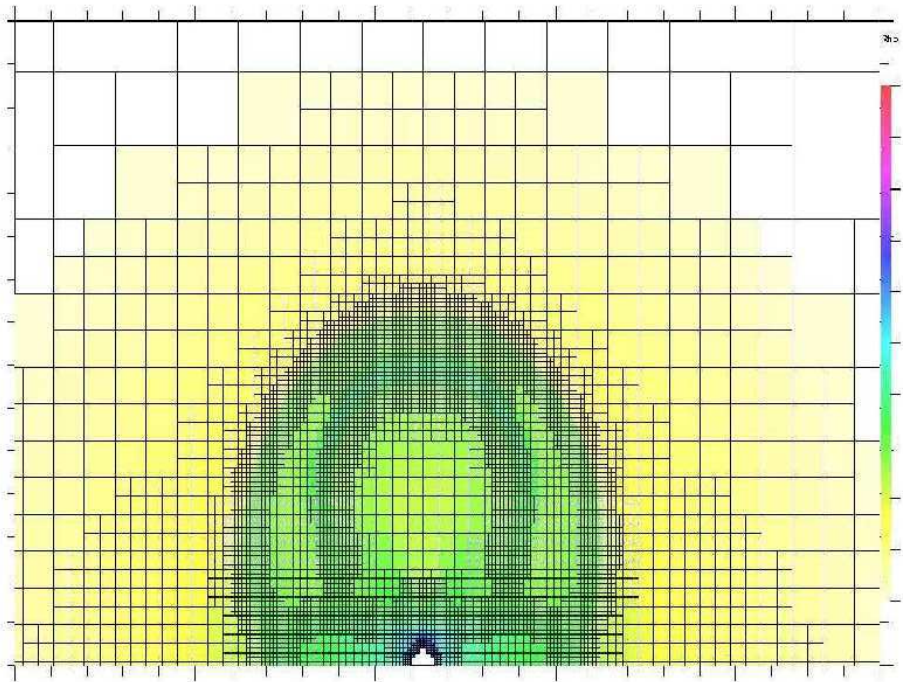}
\vskip -0.04cm
\caption{
Jet-induced model for the explosion of a C/O core (SNIc) using
a 3-D code with adaptive mesh refinement using symmetric multi-processing,
i.e. the same equations are solved by all processors but for various
sub-domains.  The total computational domain is $1.5\times 10^{11}cm$ with
a resolution of $5 \times 10^7 cm$.
 The snapshot shows the density distribution ($1.2 < log (\rho) < 7$, in CGS)
 after about 10 seconds for
the explosion of a star with 2.5 $M_\odot$ triggered by low velocity jet
($\approx 11,000 km/sec, E_{jet,total}= 1.6E50 erg$)
(\cite{k98}, H\"oflich et al. 2001).
}
\label{jet}
\end{figure}
\subsection{Domain-based parallization}

 Here, the computational domain is  subdivided and the  workload is
distributed over several CPUs. We use this approach for 3-D hydro (\cite{k98})
 and  for the spherical radiation transport in the comoving frame (Mihalas, Kunacz \& Hummer 1975).
  For the hydro, each processor solves the set of equations in a specific  spatial domain.
 The hydro code solves the compressible reactive flow equations in Eulerian frame using
an explicit Piecewise Parabolic Method (PPM) (\cite{cw84}).  For Adaptive Mesh Refinement (AMR),
it uses a Fully Threaded Tree (FTT).
   Communication between neighboring domains  is needed to exchange the results at the boundary.
The scalability is  limited  by the latency. For typical problems with an effective
resolution of $1000^3$ to $2000^3$ (see Fig. 4),
the code scales well up to 20 and 100 processors 
with an efficiency of 0.5 to 0.2 on systems with a low latency (e.g. IBM-SP3, table 1). On our
 Beowulf cluster with fast-Ethernet, the absolute peak  performance is already reached on 8 processors with an
 efficiency of $0.15$ due to the large latency of the cluster (see Table 1).
 Another example is the spherical radiation transport module for  expanding atmospheres.
The equations are discretized by ND depth and NFR frequency points. Angular resolution is achieved by
integrating along ND rays parallel to a main axis with impact parameters according to the radial grid.
  Typically, we use between 90 and 900 depth points
and $10^4$ to $10^5$ frequency points.
 Because the velocity field is monotonic, the intensity $I$ at the frequency IFR depends on the
results at the previous frequency IFR-1 making, on a first glance, the frequency space not suitable
for parallization despite the large workload which includes the integration over all rays.
 In principle, parallization could be done along individual rays. However, the number of
operations depends on the number of shells touched by a ray. As a consequence, the workload
of individual tasks will vary by a factor of ND and some
tasks would have very little work per communication. Both results in a very low efficiency.
Instead, we
distribute the workload using frequency groups, and employ load balancing as
described above. As pay-off, the boundary conditions of each frequency group depends on the neighbors and
the solution has to be iterated to obtain the correct intensity $I$ at the boundary.
 In practice,
the solution for $I$ is pretty well known from the model at a previous time step or the  NLTE iteration. Typically
 the efficiency is 30 to 50 \% even on our Beowulf cluster with a fast-Ethernet connection.

\subsection {Physical Modules}

 We found it very  useful  to employ  parallization which is based on physical modules, e.g. 
to solve the statistical  equations for all elements at a specific grid point  (see Fig.  1).
Each physical module is designed to solve a specific problem such as the equation of state,
nuclear networks, the radiation transport problem etc..
 In figure  1, we indicate the basic variable used to  distribute the computational
load.

The advantages are obvious.  Firstly,
 each of the  modules consists of complex codes (with several thousands of lines)
and the solution of the problem requires a fair amount of computations. Secondly,
only the results have to be communicated although the amount of local data can be significant 
keeping small the the communication overhead.
 E.g. for the nuclear network or the rate equations, a huge amount of atomic and nuclear data are
 required but only the resulting abundances and level populations have to be communicated.
Both these properties are the basic recipe   for good scalability and efficiency.
 On the down-side, this approach requires relatively large memory per node, typically 512MB to
1GB per node.

 For the user, this approach is beneficial too. Moderate changes are required
in course of the transition from scalar or vector code to massively parallel systems.
 This is particular true for the master/slave approach. The modular approach with well defined
interfaces allows easy code verification (see above).
 The modules can be based on well tested  tested  scalar and vector codes, and modules can be parallized
and tested one at a time.
   By using a  distributed memory model, the memory space is well separated. Data are only seen globally if they are communicated,
all but avoiding memory leaks. Of course, all programmers
keep control over all local and global variables and their names. However, by mistake,
the same variable name may be used e.g. in the hydro and the nuclear reaction network which, historically,
 are based on separate codes.

  For the data communication between various modules, we use dedicated
subroutines. E.g., for the nuclear network,  all  data relevant 
are collected by the master and communicated to a  dedicated subroutine. Subsequently, this subroutine uses calls to  the communication library (PVM or
MPI2) to transfer the data to the  dedicated routine of the Nuclear Network which  makes
the data visible to the module e.g. via COMMON blocks.

\noindent
\subsection{Some Examples}

\subsubsection{NLTE-LC calculations for Supernovae:}

As example, we want to discuss the budget for a spherical NLTE-calculation of  SNe Ia light curves.
For results, see H\"oflich (2002, this volume).
 Not all parts of our code have been  parallized and, consequently, scaling is far from ideal.
In particular, the 1-D hydro module has not yet been adopted for shared memory computers.
 However, the solution of NLTE-rate equations and the radiation transport equations are fully parallel.
 In radiation-hydro problems, the solution of the Boltzmann equation for the radiation transport and rate equations for the level population
is more CPU intensive  than the hydro 
 by factors of 50 to 100.

\begin{figure}
\vskip -2.2cm
\hskip 0.3cm \includegraphics[width=4.5cm,angle=270]{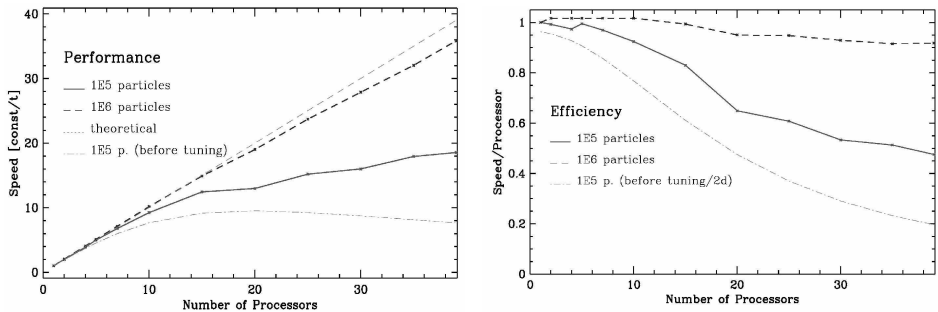}
\vskip +2.2cm
\caption{Speed increase and efficiency per processor of our Monte Carlo radiation
   transport code for various sizes of photon packages using PVM on our
  local Beowulf cluster. 
 510 radial and 120 angular points  have been used to solve the 3-D gamma-ray transport problem (see Figs.
 5 \& 6).
   The line labeled 'before tuning' corresponds
  to the same code without the use of multi-casting and without using the option for
  fast communications which avoids the format conversion of the data.}
\label{gamscal}
\end{figure}
 The calculations have been done on our 40 processor Beowulf cluster using 
spherical hydro in Lagrangian coordinates, and detailed equation of state for the atomic level
population and expansion opacities. The radiation transport is solved using the moment equations. The
Eddington factors are calculated by solving the time-independent radiation transport equation
in comoving frame. $\gamma $-ray transport is done using our Monte Carlo scheme (Fig.  1).
 For  the atomic level populations, we consider a total of 59 super-levels.
Radiation transport has been performed using 2500 frequency groups.
The envelope has been discretize by 272 depth points. To calculate the Eddington factors,
the hydro-grid has been re-mapped  to properly resolve the photosphere by 90 radial points.
a) The rate equations and the radiation transport equations are fully parallized.
  The dynamical range DR has been set to 3 for both.
Depending on the phase of the light curve, the efficiency $E(rate)$ was between 30 to 50 \%.
b) As described above, the comoving frame equations are parallized using the frequency grid.
 Formally, the problem scales almost perfect. However, we need about 2 to 5 iteration for
the boundary conditions, reducing the effective efficiency to $\approx 25  \% $.
c) The $\gamma $-ray transport
by MC is used to determine the energy deposition functions. The energy deposition varies very slowly
and hardly increases the total CPU time.
d) Finally, the 1-D hydro, moment and energy equations are tuned for shared memory systems (i.e. nodes).
The CPU-time requirements for a), b) and d) are
 $\approx $ 200, 800 and 3.8 CPU-seconds per NLTE-iteration, respectively. At each time step, we need about
5 to 10 NLTE-iteration. 2000 time steps are required for an entire light curve up to day 60.
Our LC computations take about 8 to 16 days which is rather favorable to the 110...220 days 
on a single processor system.

\begin{figure}
\hskip 0.6cm \includegraphics[width=4.6cm,angle=270]{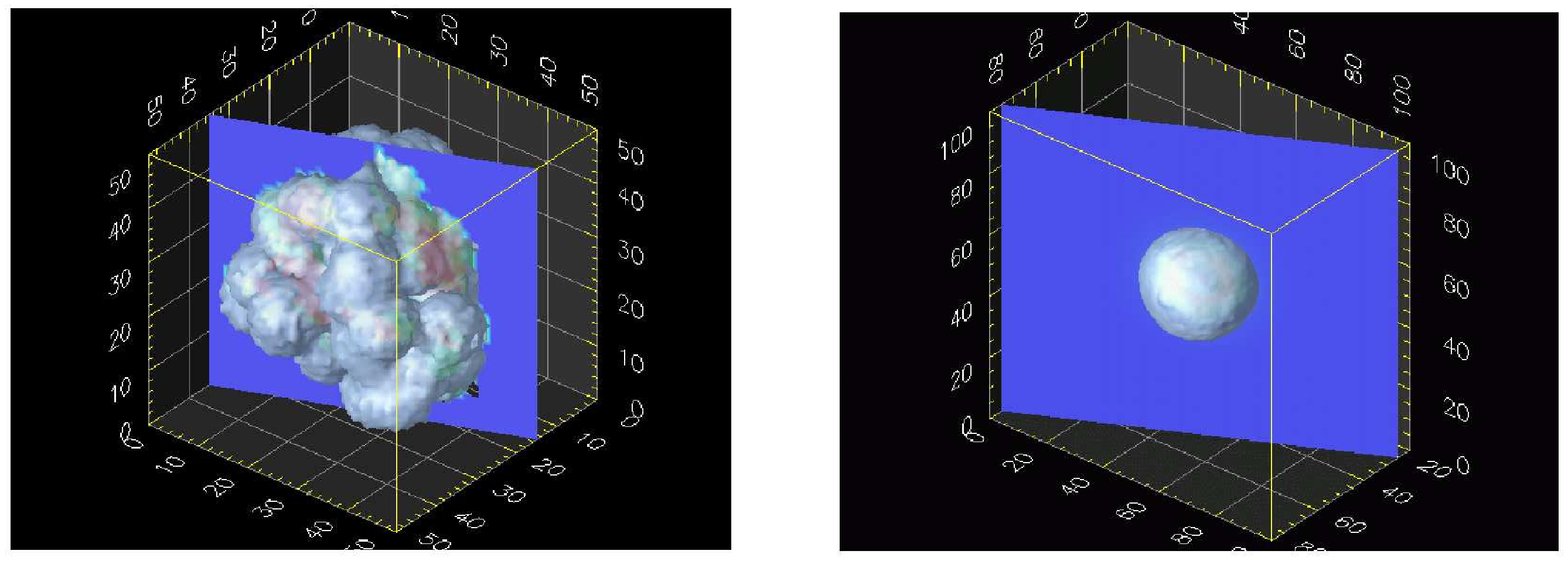}
\hskip 0.6cm \includegraphics[width=4.8cm,angle=270]{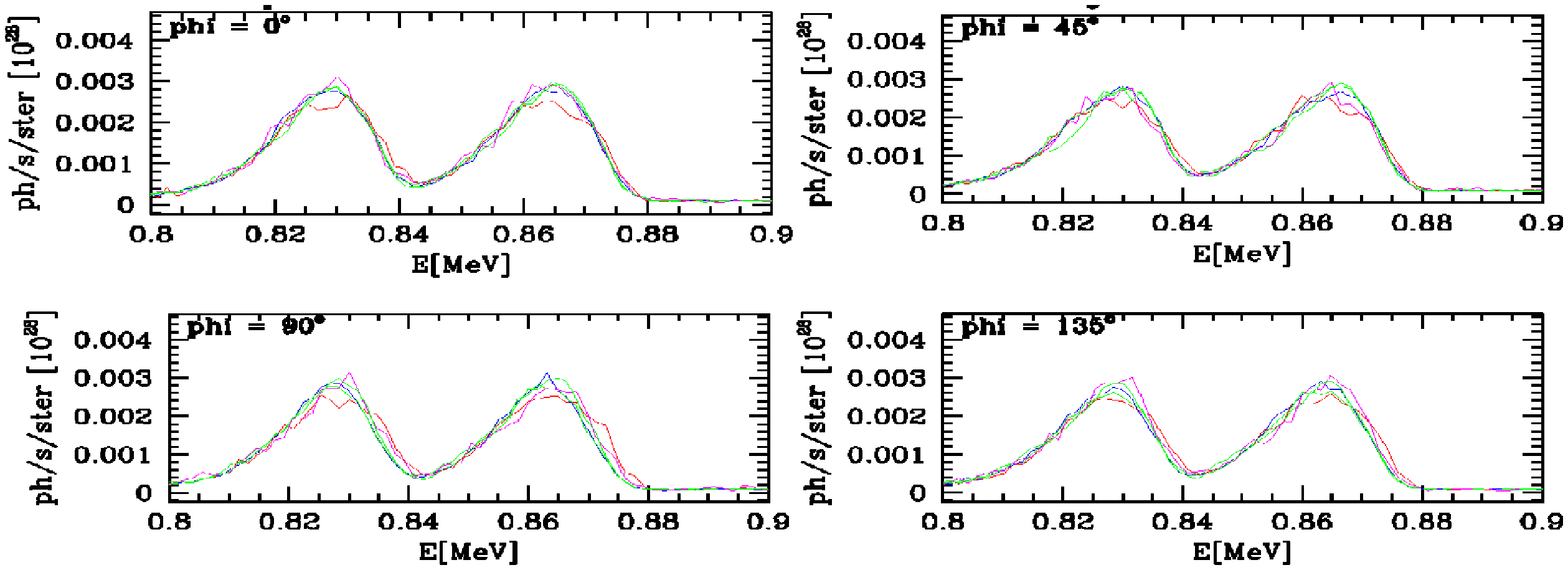}
\caption{ Massive parallel systems allow to solve 3-D radiation transport models. Here,
the energy deposition by $\gamma $-rays in a thermonuclear explosion at day 1 (left) and 23 (right).
The results have been obtained with our full 3-D  radiation transport (upper panels) (H\"oflich 2001).
 The diameter
 of the WD is normalized to 100. In the lower panels, the
  $\gamma $ spectra at day 23 are given as seen from various $\Theta $ of
  90, 60, 30 and 0$^o$ (red, blue, green, pink) and $\phi$.
 This fluctuations may be produced by instabilities in nuclear burning fronts, and they
may be seen by forthcoming $\gamma $-ray satellites such as INTEGRAL. }
\label{gam}
\end{figure}

\noindent
\subsubsection{3-D gamma ray transport:}
The final example   has been  chosen as a warning not to extrapolate scalability of a code and to
demonstrate the importance of testing and fine tuning. Monte Carlo
methods are well known for excellent scalability which, often, is called   'trivial'.
 In Fig. 5, the increase of performance and  efficiency is shows as a function of
CPUs. Apparently, scaling is nearly perfect for up to 10 processors in all cases which
include photon packages of $10^5$ and $ 10^6$ photons, and the use of  homogeneous and
heterogeneous cluster of workstations. However, with increasing number of CPUs,  the efficency drops rapidly 
for heterogeneous clusters  of
workstations. The translation of data into processor independent formats for the
communication is extremely costly and heterogeneous systems should be avoided.
 In addition, this example clearly shows the importance to use large work loads and the
influence of the 'latency'. Often the {\sl latency} of a system is more important
for the scalability  of a problem than {\sl bandwidth}.

\section{Final Discussions and Conclusions}

 Parallelization may allow to solve problems previously not feasible.
Stellar atmospheric programs and radiation hydro problems are most suitable because they use very complex
physics which, in general allow a (physical) module based approach. This enables to the use of
legacy code. Nevertheless, the efforts to change from scalar and vector-computers are 
non-trivial and time-consuming.

 Currently, three architectures are used for parallel computing: 1) shared
memory and 2) distributed memory systems, and 3) distributed systems.
 For most problems, only the first two are a valuable option.
Inhomogeneous clusters  of workstations (e.g. PCs + Suns) should be avoided.
 Often,  the scalability of a parallel code can be increased significantly by employing both
the shared and distributed memory model. Nowadays, systems become rather common which
bundle clusters of multi-processor nodes.

 Parallization tools on shared  memory systems are directive based (e.g. Open MP) or
  or done by the compiler. Typically, communication within
multi-processor nodes is sufficient fast for task distribution on a loop level.
 These systems are highly preferred for algorithm based problems such a linear equations,
matrix inversions etc.

The domain and physical module based approach is very suitable for
distributed memory systems, and load balancing can be implemented in a
'straight forward' way. In particular, parallization on a module basis
shows a very good scalability and it is easy to implement. Explicit message passing may be a bonus 
in large program packages because the memory space is well splitted which greatly  reduces the problem
of memory leaks between physical modules.
 Dynamical task allocation as a crucial feature in radiation-hydro problems which include
statistical equations  because it provides a 'fail-save' mode (see sect. 3.3).
 Fortunately, it is realized in both PVM and the new standard MPI2.
 New projects should use the communication libraries based on the MPI2 standard as soon
 as full implementations become common place.

 For high performance, it is critical to use the 'right' number of processors.
 Communication overhead may actually decrease the overall performance significantly.
 Optimization and dynamical load balancing is critical for good performance.

\begin{acknowledgments}
This research is  supported in part by  NASA Grant LSTA-98-022.
The calculations have been performed on the Beowulf cluster of the Department for Astronomy
and the HighPerformanceCenter at the University of Texas at Austin.
\end{acknowledgments}


\begin{thebibliography}{}
\bibitem[LAM]{} Byrnes et al. 2002, The LAM Foudation, http://www.lam.uc.edu
\bibitem[Collela \& Woodward 1984]{cw84}  Colella, P.; Woodward, P.R. 1984, J.Comp.Phys. {54}, 174
\bibitem[Geist et al. 1994]{geist94} Geist A., Beguelin A. Dongorra J., Jiang W., Mancheck R., Sunderam V. 1994, PVM, MIT-Press
\bibitem[Hauschild P. 2002]{Hau02} Hauschild P., 2002, this volume
\bibitem[H\"oflich, 1995]{H95} H\"oflich, P. 1995, { ApJ} {443}, 89
\bibitem[H\"oflich  1988]{H88} H\"oflich, P. 1988,   PASP {7}, 434
\bibitem[H\"oflich  2002]{H02}    H{\" o}flich  P.         2002, New Astronomy, in press \& astro-ph/0110098
on Relativistic Astrophysics, eds. Wheeler \& Martel, AIP Conference Proceedings 586, p. 459
\bibitem[H\"oflich  2002]{H02d} H\"oflich P. 2002, this volume
\bibitem[Howell et al.  2002]{HHWW01} Howell, D.~A., H{\" o}flich, P., Wang, L., \& Wheeler, J.~C.\ 2001, \apj, 556, 302
\bibitem[Khokhlov et al. 1998]{kel99}  Khokhlov A., H\"oflich P., Oran E.S., Wheeler J.C.,  P. Wang L., 1999,  ApJ 524, L107
\bibitem[Khokhlov 1998]{k98}  Khokhlov, A.M. 1998, J.Comput.Phys., 143, 519
\bibitem[Khokhlov(2001)]{khokhlov01} Khokhlov, A. 2001, ApJ , submitted \& astro-ph/0008463
\bibitem[Mihalas, Kunacz \& Hummer  1975]{MKH75} Mihalas D., Kunasz R.B., Hummer D.G. 1975
\bibitem[MPI]{mpi1995} MPI 1994,  A message passing interface standard, International Journal of Supercomputer Applications 8,
\bibitem[MPICH]{mpich2002} MPICH, 2002, A Portable Implementation of the Message Passing Interface,
 \verb"www-unix.mcs.anl.gov/mpi/mpich/"
\bibitem[MPI2]{mpi2} Message Passing Interface Forum, 1997, Documents can be found on \verb"http://www-unix.mcs.anl.gov/mpi/index.html"
\bibitem[OpenMP Architecture Review Board 2000]{}OpenMP Architecture Review Board, 2000 \verb"http://www.openmp.org"
\bibitem[Snir et al. 1995]{snir95}Snir M., Otto S., Huss-Lederman S., Walker D., Dongarra J., 1995 MIT Press and \verb"www.netlib.org/utk/papers/mpi-book/mpi-book.html"
\bibitem[Sunderam 1990]{sunderam90}Sunderam V.S. 1990, in Concurrency: Practice and Experience, p. 315, and  \verb"www.epm.ornl.gov/pvm/"
\end{thebibliography}
\end{document}